%% file: sample-sigconf.tex
\documentclass[sigconf]{acmart}

\usepackage{booktabs} 
\usepackage{multirow}
\usepackage{subcaption}
\usepackage{hyperref}
\usepackage{enumitem}
\usepackage{todonotes}
\usepackage{array}

\newcolumntype{L}[1]{>{\raggedright\arraybackslash\hspace{0pt}}m{#1}}

\setcopyright{none}

\acmYear{2018}

\acmPrice{00.00}

\input{macro}

\begin{document}
\title{The Effect of Noise on Software Engineers' Performance}

\author{Simone Romano}
\affiliation{\institution{University of Basilicata}\city{Potenza}\country{Italy}}
\email{simone.romano@unibas.it}

\author{Giuseppe Scanniello}
\affiliation{\institution{University of Basilicata}\city{Potenza}\country{Italy}}
\email{giuseppe.scanniello@unibas.it}

\author{Davide Fucci}
\affiliation{\institution{University of Hamburg}\city{Hamburg}\country{Germany}}
\email{fucci@informatik.uni-hamburg.de}

\author{Natalia Juristo}
\affiliation{\institution{Universidad Politecnica de Madrid}\city{Madrid}\country{Spain}}
\email{natalia@fi.upm.es}

\author{Burak Turhan}
\affiliation{\institution{Brunel University London}\city{London}\country{UK}}
\email{burak.turhan@brunel.ac.uk}

\renewcommand{\shortauthors}{S. Romano, G. Scanniello, D. Fucci, N. Juristo, and B. Turhan}

\begin{abstract}
	
\textit{Background:} Noise, defined as an unwanted sound, is one of the commonest factors that could affect people's performance in their daily work activities. The software engineering research community has marginally investigated the effects of noise on software engineers' performance.

\noindent
\textit{Aims:} We studied if noise affects software engineers' performance in: \textit{(i)}~comprehending functional requirements and \textit{(ii)}~fixing faults in source code.

\noindent
\textit{Method:} We conducted two  experiments with final-year undergraduate students in Computer Science. In the first experiment, we asked 55 students to comprehend functional requirements exposing them or not to noise, while in the second experiment 42 students were asked to fix faults in Java  code. 

\noindent
\textit{Results:} The participants in the second experiment, when exposed to noise, had significantly worse performance in fixing faults in source code. On the other hand, we did not observe any statistically significant difference in the first experiment. 

\noindent
\textit{Conclusions:} Fixing faults in source code seems to be more vulnerable to noise than comprehending functional requirements. 

\end{abstract}

%
%



\maketitle

\section{Introduction}\label{sec:introduction}\input{Introduction.tex}
\section{Related Work and Background}\label{sec:relatedWork}\input{RelatedWork.tex}
\section{Study Design}\label{sec:experiment}\input{Experiment.tex}

\section{Results}\label{sec:results}\input{Results.tex}

\section{Discussion} \label{sec:discussion}\input{Discussion.tex}
\section{Conclusion}\label{sec:conclusion} \input{Conclusion.tex}

\bibliographystyle{ACM-Reference-Format}
\bibliography{bibliography} 

\end{document}

%% file: macro.tex
\newcommand{\MYCOMMENT}[1]{}

\usepackage{xspace}

\newcommand{\ie}{\textit{i.e.,}\xspace}
\newcommand{\eg}{\textit{e.g.,}\xspace}

\newcommand{\etal}{\textit{et al.}\xspace}

%% file: Introduction.tex
Peopleware refers to one of the three aspects of computer technology (hardware and software are the other two). It concerns anything that has to do with the role of people in software development~\cite{constantine1995constantine}. Peopleware might cover issues related to productivity, organizational factors, workspaces, and so~ on~\cite{Acuna:2010:SPM:1941665}.

Nowadays, workspaces tend to have less privacy, with less dedicated space, which leads to noisy environment. The  reason for this trend is the cost. A ``penny'' saved on the workspace is a ``penny'' earned on the bottom line, or so the logic goes~\cite{DeMarco:2013}.  The savings of a cost-reduced workplace are attractive, but they need to be compared to the risk of performance reduction in daily working activities/tasks. 
Software companies that provide a noisy workplace are comforted by the belief that this factor does not matter~\cite{DeMarco:2013}, but noise exerts its specific influences on various forms of cognitive responses~\cite{Szalma:2011}. After all, software engineers are knowledge workers---they need to have their brain in gear to do their work---and thus their performance would be sensitive to a noisy workplace.

In this paper, we present the results of two controlled experiments, whose overarching goal was to assess if noise influences software engineers' performance. In the first experiment, we asked 55 final-year undergraduate students in Computer Science to  comprehend  functional requirements in normal conditions or exposing them to noise. The results indicated the absence of a statistically significant difference in the comprehension of functional requirements. Bearing in mind that noise could exert its influences on people's performance and these influences could be related to the task~\cite{Szalma:2011,Broadbent1978,Hancock1989}, we asked the students in the first experiment to take part in a second one (42 agreed to take part in), where we varied the kind of software engineering task. That is, participants were asked to fix faults in Java source code. The participants in the second experiment had significantly worse performances in fixing faults in source code when exposed to noise. 

The  main  contribution  of  our  paper  is  to  show  the  outcomes  of  the  first  empirical  investigation  on the effect of noise  in the comprehension of functional requirements and in the fixing of faults in Java code.  The  results suggest that there are more resource-demanding tasks and noise seems to negatively impact the performances of  software engineers when dealing with   this kind of tasks.  

\textbf{Paper Structure.} In Section~\ref{sec:relatedWork}, we present related work and background. We show the design of our investigation in Section~\ref{sec:experiment}, while the obtained results are highlighted and discussed in Section~\ref{sec:results} and Section~\ref{sec:discussion}, respectively. 
Final remarks conclude the paper.

%% file: RelatedWork.tex
In this section, we first review the literature related to our study, and then we summarize the reference theories concerning the effects of noise on individuals' performance. 

\subsection{Related Work}
The study of \textit{developers' experience}---the considerations that software developers have towards their professional activities---focuses on personal characteristics, such as feelings~\cite{Graziotin:2014cp}, motivations~\cite{Fagerholm:2012kv}, and workflow~\cite{Kuusinen:2016iq}, while neglecting the physical environment in which they operate. 
Nevertheless, a better developers' experience is thought to improve not only the job quality of software developers but also their productivity.

There have been a series of studies trying to understand how a particular mental state of software developers impacts their activities. 
\textit{Flow} is a state of high concentration that results in an absolute assimilation in the activity at hand (\eg software development)~\cite{9780062283252}. 
Disturbance from the surrounding environment in which the activity is taking place (\eg due to noise) can disrupt a state of flow causing work fragmentation and negative impact on productivity~\cite{Meyer:2017dy}. 
Alongside, Meyer \etal~\cite{Meyer:2014ih} found that developers feel productive when they do not need to switch between tasks and are not interrupted. 

Researchers have long recognized the detrimental effects of disturbance in the developers' workplace; subsequently the software engineering and HCI (Human Computer Interaction) communities have proposed different solutions to the problems related to developers' interruptibility.
Gievska \etal~\cite{Gievska:2005tp} devised an interruptibility model to mediate human interruptions by a computer. 
In a study involving 24 knowledge workers (but not software developers), they showed, through the application of their model, that reducing interruptions can increase the perceived quality of work while decreasing frustration.
For the specific case of software developers, Iqbal and Bailey~\cite{Iqbal:2007:UDM:1240624.1240732} proposed a system that would postpone possible causes of disturbance to a more apt time based on cognitive theory. 
In a study with six professionals, they showed that their approach reduces frustrations while yielding to faster reaction time.
More recently, Z{\"u}ger \etal~\cite{zuger2017reducing} developed a physical device that would signal to the surrounding environment (\eg co-workers) the best moment to disturb (or not) a developer based on her computer interaction data.
They carried out a field study with 449 participants (\ie knowledge workers, 119 of which working in software engineering-related activities), showing increased awareness about the disrupting effects of such kind of disturbance.

The studies above show that software developers' interruptibility is a topic worth investigating. However, there is only one study that focuses to some extent to noise as a source of interference with software developers' work. DeMarco and Lister related noise and other environmental factors (\eg space) to software developers' performance~\cite{DeMarco:1985}. In particular, in a study with 166 professionals working on a benchmarking task, they showed that a quiet and commodious workplace could improve productivity (\eg time to complete the task) by a factor of 2.6. There are a number of differences between the study by DeMarco and Lister and that we present in this paper. 
The main differences can be summarized as follows: we conducted two studies in controlled conditions (\eg noise was measured in our case, while in the study by DeMarco and Lister participants provided their perception on the noise level in their workplace) on two kinds of software engineering tasks  (SE task/s from here onwards) and we quantitatively assessed the effect of noise on performances (\ie achieved comprehension of functional requirements and capability to fix faults in source code). An additional difference is related to the programming language of the used experimental object (COBOL vs Java).


\subsection{Background}\label{sec:background}\input{BackGround.tex} 

%% file: BackGround.tex
We highlight the reference theories defined to explain and predict noise effects on individuals' performance.

\subsubsection{Arousal Theory} 

To explain noise effects, Broadbent~\cite{Broadbent1978} invoked an arousal induced attentional narrowing mechanism in the individuals. In Broadbent's theory, noise increases arousal of an individual, which decreases his/her breadth of attention. At relatively lower levels of arousal, individuals exclude task-irrelevant cues, and thus the attentional narrowing facilitates performance. However, beyond a certain arousal ``optimal'' level,  individuals' performance is impaired because increases in arousal might cause increased narrowing so that task-relevant cues are excluded.
Arousal theory~\cite{Broadbent1978} predicts that more demanding tasks should have lower levels of optimum arousal, and thus these tasks should yield the greatest performance decrements in the presence of noise. Hence, cognitive tasks should suffer greater magnitudes of performance impairment with respect to less demanding tasks (\eg psychomotor ones). The intensity of noise and duration of noise exposure influence the arousal levels; \ie higher intensities and longer durations should cause greater negative effects on performance. Concerning noise schedule, intermittent noise should impair performance more than continuous one. In summary, in the Broadbent's theory noise effects should vary according to the kind of task and the noise intensity, duration, and schedule.



\subsubsection{Composite Theory}
Poulton's~\cite{Poulton1979}  theory predicts that noise effects should degrade individuals' performance only for those conditions in which inner speech\footnote{Also referred to as verbal thinking, inner speaking, covert self-talk, internal monologue, and internal dialog. Inner speech is thinking in words and also refers to the semi-constant internal monologue some individuals have with themselves at either conscious or semi-conscious level~\cite{10.3389/fpsyg.2015.01758}.} is masked. The gains in individuals' performance in continuous noise early in the task occur because the increase in arousal compensates for the detrimental effects of masking. However, with time on task, arousal decreases and thus masking effects dominate. The way in which arousal affects performances is different between composite and arousal theories. Noise intensity could also affect performance. In summary, noise effects should be similar across task and noise kind, but moderating effects are expected for intensity, duration, and schedule.

%
%


\begin{table*}[t]
	\centering
	\caption{Summary of the experiments.}
	\vspace{-0.4cm}
		\label{tab:summary}
	\resizebox{0.7\linewidth}{!}{
	\begin{tabular}{lll}
		\hline
		Characteristic & \multicolumn{1}{c}{Exp1}                                     & \multicolumn{1}{c}{Exp2}                      \\ \hline \hline
		Schedule & 11:30 on 2016/12/12 & 11:30 on 2017/31/1 \\
		Kind of SE task                    & Comprehension of functional requirements & Fault fixing in source code             \\
		SE task duration                   & 30 minutes                               & 60 minutes                \\
		Experimental objects            & M-Shop and Theater           &  LaTazza and AveCalc \\
		Participants                    & 55 Undergraduate students in Computer Science                & 42 Undergraduate students in Computer Science \\
		Group1/Group2 size              & 28/27                                    & 21/21                     \\
		Experiment design               & AB/BA crossover design                   & AB/BA crossover design    \\
		Investigated RQ                 & RQ1                                      & RQ2                      \\ \hline
	\end{tabular}
	}
\end{table*}

\subsubsection{Maximal Adaptability Theory}
In the maximal adaptability model~\cite{Hancock1989}, stress (noise is a source of stress) can be accounted for in three loci. \textit{Input} represents all objective environmental and task factors that affect performance (\eg noise), \textit{adaptation} concerns the capacity of the individual to cope with demands intrinsic to an environment (\eg physiological coping responses), and \textit{output} refers to the individual's response about the task environment. The output of a task depends on the characteristics of individuals, and it might be directly affected by noise. As for adaptation, noise can impair the capacity through the masking or distortion of task-relevant auditory information. 
According to the maximal adaptability theory, individuals can adapt to a quite broad range of stress magnitudes. However, there is a threshold of dynamic instability in which adaptation fails, and thus performance decreases. 
The maximal adaptability theory predicts that performance on more resource-demanding cognitive tasks, in case of noise, should be more impaired than performance on motor or perceptual tasks. For higher noise intensities and longer noise durations, there should be a greater performance impairment. Concerning the kind of noise, speech noise should be more disruptive than non-speech, especially in cognitive tasks. Also, noise schedule could affect performance. In summary, the maximal adaptability theory predicts that noise effects should vary as a function of task and noise kind, and schedule, duration, and intensity.

\subsubsection{Empirical Evidences from Noise Effects on Performance}

Arousal, composite, and maximal adaptability theories predict similar results on noise effects for certain  variables (\eg noise intensity, duration, and schedule), but different results for others (\eg kind of task and noise). Szalma and Hancock~\cite{Szalma:2011} have recently conducted a meta-analysis on noise effects on individuals'  performances. Results confirm only in part the predictions of the three reference theories, and in some cases are inconsistent with such predictions. That is, the meta-analytic results confirm that noise effects varied as a function of the kind of noise and task, and noise intensity, duration, and schedule. However, Szalma and Hancock reported that shorter durations have greater detrimental effects on performance than longer durations. 


%% file: Experiment.tex
In Table~\ref{tab:summary}, we summarize the main characteristics of our study, which comprises two controlled experiments. We refer to these experiments as Exp1 and Exp2 (see the second and third columns of Table~\ref{tab:summary}), respectively. Exp1 was conducted on 2016/12/12, while Exp2 on 2017/31/1. 
In both experiments, we investigated whether noise affects performance when carrying out some SE tasks.
The participants in Exp1 had to comprehend functional requirements of two software systems (\ie M-Shop and Theater as shown in the fourth row of Table~\ref{tab:summary}) exposing them or not to noise. Similarly, the participants in Exp2 had to fix faults in two Java programs (\ie LaTazza and  AveCalc   as shown in  Table~\ref{tab:summary}).


To perform our study, we followed the guidelines by Juristo and Moreno~\cite{juristo}, and Wohlin \etal~\cite{wohlin12}. We present the design of our study according to Jedlitschka \etal's guidelines~\cite{Jedlitschka2008}.  

\subsection{Goals}
The goal of our study, according to the Goal Question Metrics (GQM) template~\cite{basili94}, is:

\vspace{0.05 cm}
\noindent
\textbf{Analyze} noise \textbf{for the purpose of} evaluating its effect \textbf{with respect to} the performances in comprehending functional requirements and in fixing faults in Java source code \textbf{from the point of view of} the researcher \textbf{in the context of} final-year undergraduate students in Computer Science.
\vspace{0.05 cm}

\noindent
Therefore, we investigated the following research questions:
\begin{description}[leftmargin=*]
	\item[RQ1] Does noise worsen software engineers' performance in comprehending functional requirements?
	\item[RQ2] Does noise worsen software engineers' performance in fixing faults in source code? 
\end{description}

\subsection{Experimental Units}
The participants in our experiments were final-year undergraduate students in Computer Science at the University of Basilicata. They had programming experience in Java and basic knowledge of software design, development, and testing.
The participants attended the Software Engineering (SE) course in which we conducted both experiments as optional laboratory exercises. This course taught advanced elements concerning:  software development processes, requirements specification, software design, maintenance, and testing. During the SE course, the participants carried out  homework and classwork on requirements specification and bug fixing to increase their technical maturity on the topics covered in the course.    

To encourage the participation in our study, we rewarded the students who took part in Exp1 with a bonus, \ie one point on their final mark in the SE course. 
Not all the participants in Exp1 took part in Exp2; a subset of the participants in the first experiment took part in the second one too. Students who took part in both experiments received two points of bonus. We communicated to the participants that their performance in the experiments would not affect their grade, and that the collected data would be shared anonymously and used for research purposes only. The participation in both experiments was on voluntary basis (\ie in no case we obliged students to participate in the experiments).

\subsection{Experimental Material} \label{sec:expMaterial}
As for Exp1, the chosen experimental objects where:
\begin{itemize}[leftmargin=*]
	\item \textbf{M-Shop}\textemdash A system for managing the sales in a music shop.
	\item \textbf{Theater}\textemdash A system for managing the reservation of tickets in a~theater.
\end{itemize}
For each of these systems, one functional requirement with the corresponding models (\ie functional model, analysis object model, and dynamic model) was selected from its requirements analysis specification.\footnote{The use of incomplete documentation and of a subset of the entire software system on which a maintenance operation impacts is quite common in the software industry~\cite{bruegge2003}.  Examples are when only part of the documentation exists (\eg in lean development processes), is up to date, or is useful to perform a given SE task.} In particular, the selected functional requirement for M-Shop was ``Search Album by Singer,'' while for Theater was ``Buy Theater Ticket.'' We chose these systems, and we selected these functional requirements because they were previously used in a family of controlled experiments to assess whether the comprehension of functional requirements was influenced by the use of dynamic models (represented through UML sequence diagrams)~\cite{Abrahao2013}. The authors, who conducted this family of experiments, administered the participants in the control group with the functional model and analysis object model associated with the selected functional requirement. The participants in the treatment group were administered with the same models as the control group plus the dynamic models (\ie UML sequence diagrams). To evaluate the comprehension of functional requirements, the authors asked the participants to fill out comprehension questionnaires. Assessing comprehension of software artifacts (\eg models or source code) through questionnaires is common in SE experiments (\eg \cite{KamstiesKR03,Ricca2014}). We exploited in Exp1 the experimental material, \ie models and questionnaires, the authors~\cite{Abrahao2013} made available on the web and they administered to the participants in the treatment group. This design choice should not affect the results because the participants who accomplished the task in noise conditions were provided with the same material as the participants who accomplished the task in normal conditions.

As for Exp2, we chose the following experimental objects:\\
	\textbf{LaTazza}\textemdash A Java desktop application for managing the sale and
	the supply of small-bags of beverages (coffee, arabica coffee, tea) for a coffee maker. Its source code had 18 classes and 1,215 LOC (\ie Lines of Code).\\
	\textbf{AveCalc}\textemdash A Java desktop application for managing the exams of a student during its university career. Its source code had 33 classes and 1,388 LOC.\\
LaTazza and AveCalc were used in other empirical studies (\eg~\cite{ricca2008, Scanniello2017}). In particular, we exploited the source code of the experimental objects used in the three experiments by Ricca \etal~\cite{ricca2008} (the source code they administered to the control group was the same as the treatment group). To assess the performance in fixing faults in source code, we provided the participants with bug reports and asked them to fix the faults that such bug reports described. We exploited the bug reports Scanniello \etal~\cite{Scanniello2017} defined on the experimental objects by Ricca \etal~\cite{ricca2008} and then used in their family of controlled experiments (the bug reports administered to participants were the same in both treatment and control groups). Similar to Exp1, we used the experimental material (\ie source code and bug reports) the authors~\cite{ricca2008,Scanniello2017} made available on the web. It is worth mentioning that assessing the performance in fixing faults in source code with a different instrumentation tool (\eg picking the correct fix for a bug from a multiple-choice question) would have decreased the realism of the fault fixing tasks. Conversely, the used instrumentation tool allowed reducing threats to external validity.

We used  experimental materials defined by different researchers to mitigate experimenters' expectancies biases. We made both experimental material and raw data available on the web.\footnote{\href{http://www2.unibas.it/sromano/downloads/NoiseExpsReplicationPackage.zip}{www2.unibas.it/sromano/downloads/NoiseExpsReplicationPackage.zip}}



\begin{figure}[t]
	\centering
	\resizebox{0.8\columnwidth}{!}{%
		\begin{tabular}{|L{0.25cm}p{7.5cm}|}
			\hline
			Q5. & Based on the furnished models, the primary actor can: (Mark the right answer/s) \\ 
			& $\Box$ Select any album\\
			& $\Box$ Modify an album\\
			& $\Box$ Get the available copies for an album \\
			& $\Box$ See the details of an album\\ \hline                           
		\end{tabular}
	}
	\vspace{-0.3cm}
	
	\caption{A sample comprehension question for M-Shop.}\label{fig:question}
\end{figure}

\subsection{Tasks}\label{sec:tasks}
The participants in Exp1 had to perform the following tasks:
\begin{enumerate}[leftmargin=*]
	\item \textbf{Comprehension task 1}\textemdash We provided each participant with the models 
	associated with the functional requirement ``Search Album by Singer'' of M-Shop. To evaluate the comprehension of such a requirement, we asked the participants to fill out a comprehension questionnaire consisting of 11 closed-ended questions. Each question admitted one or more right answers. In Figure~\ref{fig:question}, we report a sample question of the comprehension questionnaire of M-Shop, which admitted two right answers: ``Get the available copies for an album'' and ``See the details of an album.''
	\item \textbf{Comprehension task 2}\textemdash We asked the participants to perform the same task as the previous one, but the experimental object was Theater. In particular, we gave each participant the models associated with the functional requirement ``Buy Theater Ticket.'' Then, we asked to fill out a comprehension questionnaire similar (\eg it comprised 11 closed-ended questions) to that used in the previous~task. 
\end{enumerate}

\begin{figure}[t]
	\centering
	\resizebox{0.85 \columnwidth}{!}{%
	\begin{tabular}{|l|p{7cm}|} \hline
		\multicolumn{2}{|l|}{Start Time (hh:mm): \phantom{iiiiiiiiiiiiiiiiiiii} End Time (hh:mm): } \\ \hline	
		ID          & 2                                                                                                                                                                                                           \\ \hline
		Title       & Wrong product selection                                                                                                                                                                                     \\ \hline
		Description & If you select ``supply of small bags'' and then ``Coffee'' to buy a supply of coffee, arabica coffee is wrongly bought instead of coffee. The problem does not occur when buying a supply of arabica coffee. \\ \hline
	\end{tabular}
    }
     \vspace{-0.3cm}
    \caption{A sample bug report for LaTazza.}\label{fig:bugreport}
    \vspace{-0.4cm}
\end{figure}

As for Exp2, the participants had to perform the following tasks:
\begin{enumerate}[leftmargin=*]
	\item \textbf{Bug fixing task 1}\textemdash We provided the participants with the source code (no test cases were given) and mission (\ie problem statement) of LaTazza. We also gave them six bug reports, one for each fault the participants had to fix in the source code of LaTazza. A bug report contained the ID of the corresponding bug, as well as a title and a description. In Figure~\ref{fig:bugreport}, we show a sample bug report for LaTazza. When a participant fixed a fault, he/she had to indicate the portion of source code he/she modified to fix such a fault, \ie his/her patch. In particular, the participant had to insert two Java comments to delimit the modified code: \texttt{/*patch~\textless fault\_ID\textgreater~start*/} just before the patch, and \texttt{/*patch~end*/} at the end of the patch. This procedure is the same as used in~\cite{Scanniello2017}, where further details on the bug reports and the faults are available and that we have not reported here for space reason and scant relevance. The participants had to also write down when they start tackling each bug and when it was fixed (see Figure~\ref{fig:bugreport}). 
			
	\item \textbf{Bug fixing task 2}\textemdash We asked the participants to perform the same task as the previous one but on AveCalc. Therefore, we provided them with the source code and mission of AveCalc together with six bug reports. To indicate the patches, the participants had to use Java comments as done in the previous~task.
	 
\end{enumerate}	

\subsection{Hypotheses, Parameters, and Variables}
The participants who had to perform the tasks (\ie comprehension of functional requirements or bug fixing in source code) in the presence of noise comprised the 
treatment group;
while those who worked on the tasks in normal conditions (\ie they were not exposed to noise) comprised the 
control group. Therefore, in each experiment \textit{Condition} is the main independent variable (also known as main factor or manipulated factor or
main explanatory variable). Condition is a nominal variable that assumes two values: NOISE (\ie participants working in noise conditions) and NORMAL (\ie participants working in normal conditions).

To quantify software engineers' performance in comprehending functional requirements, we evaluated participants' answers to each comprehension questionnaire by using two strategies. The first one was an information retrieval-based strategy~\cite{Manning08}. It consists in computing precision ($P_{c}$) and recall ($R_{c}$) of the answers given by the participant $s$ as follows:
\begin{equation*}
\textstyle P_{c}=\frac{ \sum_{i=1}^n | answers_{s,i}  \cap  oracle_i|} { |\sum_{i=1}^n answers_{s,i}|}
\phantom{i}   
{\textstyle R_{c}}=\frac{ \sum_{i=1}^n | answers_{s,i}  \cap  oracle_i|} { |\sum_{i=1}^n oracle_i|}
\end{equation*}

\noindent
where $answers_{s,i}$ is the set of answers the participant $s$ provided for the question $i$, while $oracle_i$ is the set of correct expected answers (\ie the oracle) for the question $i$. We indicate with $n$ the number of questions (\ie 11 for both M-Shop and Theater). From a practical perspective, $P_{c}$ and $R_{c}$ estimate correctness and completenesses of the given answers. To get a single measure that represents a trade-off between correctness and completenesses, we compute the balanced F-measure~\cite{Manning08} between $P_{c}$ and $R_{c}$ as follows:
\begin{equation*}\label{eq:F1}
{\textstyle F_{c}}=\frac{ 2 * P_{c} * R_{c}} {P_{c} + R_{c}}
\end{equation*} 

\noindent
This metric assumes values in the interval $[0,1]$, where 1 means that the participant $s$ answered very good the questions of the comprehension questionnaire, so assuming that his/her comprehension of functional requirements was very good. Conversely, a value close to 0 means that he/she answered very bad (\ie functional requirements were scarcely comprehended).
This information retrieval-based strategy is the same as used in~\cite{Abrahao2013}. The second strategy we adopted to quantify performance in comprehending functional requirements was inspired to that used by Kamsties \etal~\cite{KamstiesKR03}.
In particular, for each participant $s$, we computed the variable $count$ as follows:
\begin{equation*}
count_{s,i} = \bigg \{ \begin{array}{rl}
1 & \textstyle if~answers_{s,i} = oracle_{i} \\
0 & \textstyle otherwise \\
\end{array}
\end{equation*}

\noindent
$count$ assumes 1 as the value if and only if the set of answers provided by the participant $s$ to the question $i$ corresponds to the oracle for the question. We quantified performance in comprehending functional requirements by means of the following metric:  
\begin{equation*} \label{Avg}
Avg=\frac {\sum_{i=1}^n count_{s,i}}{n}
\end{equation*} 

\noindent
where $n$ is the number of questions. $Avg$ assumes values in the interval [0,1], where 1 is the best possible value. In other words, the higher the $Avg$ value, the better the comprehension of functional requirements is. Unlike $F_c$, $Avg$ does not take into account partial answers. The use of two metrics (\ie $F_c$ and $Avg$) to assess the comprehension of functional requirements allowed us to mitigate possible construct validity threats (\ie mono-method bias)~\cite{wohlin12}.

To measure software engineers' performance in fixing faults in source code, we relied on the information retrieval-based strategy used in~\cite{Scanniello2017}. We estimated correctness and completeness of the faults the participant $s$ fixed by means of the precision ($P_{f}$) and recall ($R_{f}$) measures, respectively:
\begin{equation*}  
{\textstyle P_{f}}=\frac{\#~faults~correctly~fixed_s} {\#~faults~fixed_s}
\phantom{iii}   
{\textstyle R_{f}}=\frac{\#~faults~correctly~fixed_s} {\#~faults~present}
\end{equation*} 

\noindent
where $\#~faults~fixed_s$ is the count of bug the participant $s$ fixed both correctly and incorrectly. We knew this information because each participant had to delimit his/her patch when fixing a given fault (see Section~\ref{sec:tasks}). To get $\#~faults~correctly~fixed_s$, we first inspected the patches provided by the participant $s$, then we run an acceptance test suite for each bug. If this suite passed, the corresponding bug was correctly fixed. $\#~faults~present$ is the number of bug the participants had to fix (\ie six for both LaTazza and AveCalc).  

\begin{figure*}[t]
	\centering
	\resizebox{0.92\linewidth}{!}{%
		\includegraphics{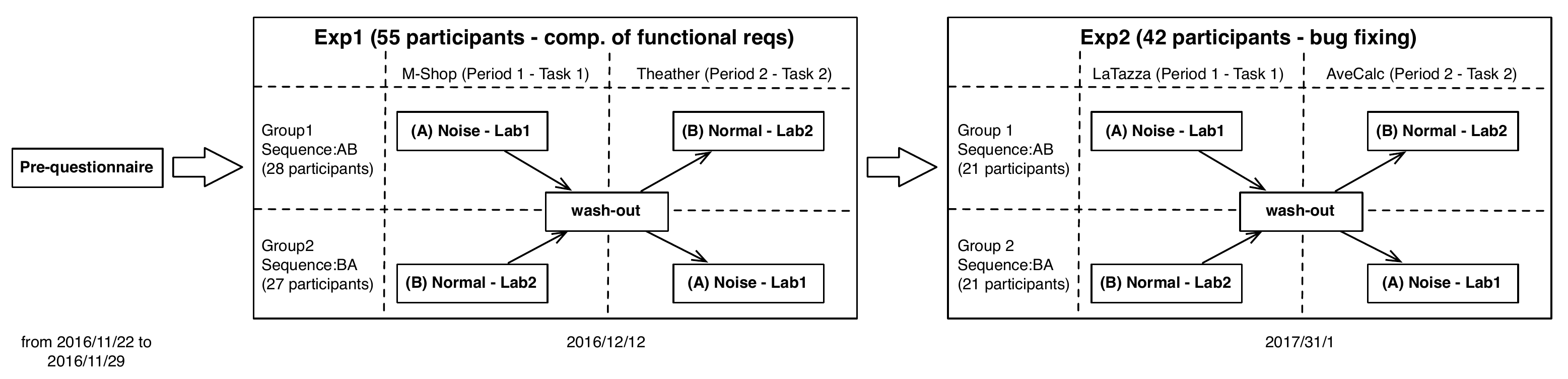}
	}
	\vspace{-0.5cm}
	\caption{Summary of the design.}\label{fig:summary}
		\vspace{-0.5cm}
\end{figure*}

To get a trade-off between correctness and completeness of fixed faults, and thus estimate performance in fixing fault in source code, we computed the balanced F-measure as follows:
\begin{equation*}
{\textstyle F_f}=\frac{ 2 * P_{f} * R_{f}} {P_{f} + R_{f}}
 \end{equation*}
A value for $F_f$ close to 0 means that the performance in fixing faults was very bad (\ie existing bugs were, for the most part, either not fixed or incorrectly fixed), while a value close to 1 means that the performance in fixing faults was very good.

According to our research questions, we formulated the following null-hypotheses:

\begin{description}[leftmargin=*]
	\item[Hn1] Noise \emph{does not significantly affect} software engineers' performance in comprehending functional requirements.
	\item[Hn2] Noise \emph{does not significantly affect} software engineers' performance in fixing faults in source code.
\end{description}
Given the literature regarding the effect of noise  (\eg \cite{Broadbent1978,Poulton1979,Hancock1989,Szalma:2011}), we expect that noise might have detrimental effect on software engineers' performance. Therefore, previous knowledge allows us to formulate the one-tailed alternative hypotheses: \textit{noise significantly and negatively affects software engineers' performance in [comprehending functional requirements $|$ fixing faults in source~code].}

%

\subsection{Experiment Design} \label{sec:design}

In Figure~\ref{fig:summary}, we graphically describe the overall design of our empirical investigation. Before Exp1 took place, the participants had to fill out a pre-questionnaire to gather their demographic information. We used Google Forms to create this questionnaire and left the participants the possibility to fill it out from 2016/11/22 to 2016/11/29. The gathered information on the participants allowed us to characterize the context of our study better. 


For both experiments, the used design was an AB/BA crossover. AB/BA crossover designs have two treatments (\ie A and B) and two periods (\ie the times at which each treatment is applied). Participants are split into two experimental groups and administered with every treatment only once. In AB/BA crossover designs, the groups are the sequences, \ie the order in which treatments are administered to participants~\cite{Vegas:2016}.

Figure~\ref{fig:summary} also shows how the AB/BA crossover design was applied to both Exp1 and Exp2. 
In our case, the treatments A and B were NOISE and NORMAL, respectively. The participants were randomly assigned to the experimental groups:\\
\textbf{Group1}\textemdash It comprises participants assigned to the sequence AB (\ie NOISE is applied in the first period, while NORMAL in the second). \\
\textbf{Group2}\textemdash It comprises participants assigned to the sequence BA (\ie NORMAL is applied in the first period, while NOISE in the second).\\
As for Exp1, the number of participants in Group1 and Group2 was 28 and 27, respectively; while 21 and 21 in Exp2. Independently from the treatment, the participants in Exp1 performed comprehension task 1 (on M-Shop) in the first period, while comprehension task 2 (on Theater) in the second. The participants in Exp2 carried out bug fixing task 1 (on LaTazza) in the first period, while bug fixing task 2 (on AveCalc) in the second. 
 Group1 and Group2 carried out each task (\eg comprehension task 1) at the same time. 

Comprehension tasks lasted 30 minutes each, while bug fixing tasks lasted one hour each. 
We fixed the time to accomplish the tasks on the basis of the experimental data reported in~\cite{Abrahao2013, Scanniello2017}. The use of tasks with different durations allowed studying the effect of  duration on performance.
The strategy we used to define performance (in both the experiments) was \textit{time fixed}, namely the number of successful steps within the time limit defines performance~\cite{BergersenSD14}. 

NOISE  was always applied in the Lab1 research laboratory, while NORMAL always in Lab2 (see~Figure~\ref{fig:summary}). We took this design decision to mitigate possible conclusion validity threats related to the experimental settings.  

We added a 30-minute wash-out period between Period~1 and Period~2 (see Figure~\ref{fig:summary}). The rationale behind the use of wash-out periods (as suggested in medicine studies) is to leave sufficient time for the effect of a treatment to recede completely, and thus possibly neutralize carryover effects.\footnote{Carryover is an internal validity threat of crossover designs. It occurs when a treatment is administered before the effect of another previously administered treatment has completely receded~\cite{Vegas:2016}.} However, it can be difficult to determine how long a wash-out period should be to recede completely the effect of a treatment~\cite{Dean:2015}. That is, the required length of wash-out period for a given treatment is usually unknown before the study takes place~\cite{Gallin:2007}.
Therefore, carryover effects could also occur in the presence of wash-out periods if they are not long enough. Therefore, we analyzed whether carryover effects were present in our experiments (see Section~\ref{sec:analysisProc}).   

\subsection{Experiment Setting} \label{sec:validityConsiderations}
The Directive 2003/10/EC\footnote{\href{http://eur-lex.europa.eu/legal-content/EN/TXT/?uri=CELEX:02003L0010-20081211}{eur-lex.europa.eu/legal-content/EN/TXT/?uri=CELEX:02003L0010-20081211}} of the European Parliament lays down minimum health and safety requirements regarding the exposure of workers to the risks arising from noise. In particular, if the noise exposure level\footnote{It is the time-weighted average of the noise exposure levels.} ($L_{EX}$) is greater than or equal to 85~dB, workers shall wear individual hearing protectors; thus values that match or exceed such a limit are considered harmful to health. The participants administered with the NOISE treatment had to accomplish the  tasks with a value of $L_{EX}$  (\ie 82~dB) close but inferior to the limit mentioned above. Thus individual hearing protectors were not required. 
The kind of noise was speech because it is the major type of practical distractive noise in workplaces with open-office plans~\cite{Szalma:2011,Maxwell}. 
As for the participants working in normal conditions, they performed the experimental tasks in a laboratory (\ie Lab2) far from road traffic and other sources of noise (the measured $L_{EX}$ value was equal to 42~dB, namely a common value of noise level for quiet office workplaces~\cite{Maxwell}). We chose the exposure value for the NOISE treatment on the basis of the results from the study by Szalma and Hancock:  low-intensity noise is more debilitating than high-intensity noise for those studies in which quiet is the control condition~\cite{Szalma:2011}.  Therefore, if we found a significant effect of noise, we would expect an higher effect of noise for lower $L_{EX}$ values.

The NOISE treatment was administered in a laboratory (\ie Lab1) equipped with a sound system. A (ceiling) speaker was present on each workstation used by each participant. The distance between each speaker and each workstation was always the same. Before the experiments took place, for any workstation we measured the noise level chosen for the experimental setting with a sound meter. In particular, we placed the sound meter 45 cm above the desk. Such a measurement allowed us to grasp, with some approximation,\footnote{This is an approximated measurement because different participants' posture and height could slightly affect the noise levels the participants heard.} what a participant would have heard during the tasks.  We measured the noise levels also during the experiments. Lab1 was far from road traffic (\ie the most dominant source of environmental noise~\cite{european2014noise}) and other sources of noise that might have interfered with the noise reproduced by the sound system.


Furniture (\eg the kind of desk or chair) in Lab1 was the same as in Lab2. These laboratories were equipped with the same PCs (\ie same hardware and software configurations) and the environmental configurations were similar (\eg similar light level) as well as the design (\ie open-office). This is because we wanted to mitigate environmental factors that might affect results in an unexpected~way. Similar to Lab1, Lab2 was far from sources of noise.

\subsection{Analysis Procedure}\label{sec:analysisProc}
We used the following analysis procedure for each experiment and each dependent variable. According to Wellek and Blettner's study~\cite{DAE124835}, we perform a pre-test to check the presence of a carryover effect. Let $X_{1i}$ and $X_{2i}$ be the dependent variable values (\eg $F_c$) for the participant $i$ of Group1 (\ie sequence AB) in the first and second periods, respectively. Similarly, let $Y_{1j}$ and $Y_{2j}$ be the dependent variable values for the participant $j$ of Group2 (\ie sequence BA) in the first and second periods, respectively. For each participant $i$ of Group1 and each participant $j$ of Group2, we compute the within-participant sums of the dependent variable values in both periods as follows: $C_i(X) = X_{1i} + X_{2i}$ and $C_j(Y) = Y_{1j} + Y_{2j}$. 
If data are normally distributed, we run an unpaired two-sided t-test to verify the null-hypothesis: are the expected mean values of within-participant sums  the same~\cite{DAE124835}? In case data are not normally distributed, we run a two-sided Mann-Whitney U test~\cite{Conover1998} (also known as Wilcoxon rank-sum test). 
The Mann-Whitney U test is a non-parametric alternative to the t-test~\cite{wohlin12}. If the t-test/Mann-Whitney U test does not reject the null-hypothesis, the carryover effect is not statistically significant. 

If carryover effect is not statistically significant, we performed the following steps: 
\begin{enumerate}[leftmargin=*]
\item We computed descriptive statistics and build boxplots for each dependent variable.
\item We tested whether the effect of noise was statistically significant on the dependent variable. To this end, for each participant $i$ of Group1 and each participant $j$ of Group2, we computed the within-participant differences of the dependent variable values in both periods~\cite{DAE124835}: $D_i(X) = X_{1i} - X_{2i}$ and $D_j(Y) = Y_{1j} - Y_{2j}$. Then, if the data were normally distributed, we run an unpaired two-sided t-test, where the tested null-hypothesis is: are the expected mean values of within-participant differences  the same~\cite{DAE124835}? We run a two-sided Mann-Whitney U test, if  data are not normally distributed.
\end{enumerate}

If carryover effect is statistically significant for a given  depended variable, we discarded the second period~\cite{Vegas:2016}. That is, we  analyzed the first period  as follows:
\begin{enumerate}[leftmargin=*]
\item We computed descriptive statistics and build boxplots.
\item We tested if the effect of noise was statistically significant on the dependent variable values by means of an unpaired two-sided t-test, if the data were normally distributed, or a two-sided Mann-Whitney U test otherwise.
\end{enumerate}

To check normality of data, we used the Shapiro-Wilk test~\cite{shapiro65} (Shapiro test, from here onwards).

As it is customary with tests of statistical significance, we accept a probability of 5\% of
committing Type-I error (\ie $\alpha= 0.05$). 

Since 42 out of 55 participants took part in both the experiments, we also studied the performances of the  participants in Exp1 who also participated in Exp2. We refer to this subset as Exp1*, where  Group1 and Group2 comprised 21 participants each. The  analysis procedure for Exp1* was the same as that used for Exp1 and Exp2. The goal of this further analysis was to see if the results from Exp1 are confirmed or not by those from Exp1*. In case of confirmation, we can state that the results from Exp1 were not due to any characteristic of the participants who did not take part in~Exp2.

%% file: Results.tex
\begin{table}[t]
	\centering
	\caption{Results from the pre-test to check the presence of carryover effects (in bold p-values less than $\boldsymbol \alpha$).}
	\label{tab:pretestRes}
	\resizebox{0.48\columnwidth}{!}{
	\begin{tabular}{llc}
		\hline
		Experiment            & Variable & p-value \\ \hline \hline
		\multirow{2}{*}{Exp1} & $F_c$    & 0.5179          \\
		& $Avg$    & 0.4312          \\ \hline
		\multirow{2}{*}{Exp1*}& $F_c$    & 0.2432          \\   
		& $Avg$    & 0.1809          \\ \hline            
		Exp2                  & $F_f$    & \textbf{0.0358} \\ \hline
		\end{tabular}
}
\end{table}

In this section, we report the results of our analysis. 

\subsection{Carryover Analysis} \label{sec:carryover}
The Shapiro test suggested that the data were normally distributed in each experiment and for each dependent variable (\ie p-values were greater than $\alpha$), thus we applied a t-test to check the presence of a carryover effect. The obtained p-values are reported in Table~\ref{tab:pretestRes}.
The results indicate that the carryover effect is not statistically significant for any dependent variable in Exp1 and Exp1*. The carryover effect is statistically significant for $F_f$ in Exp2 (p-value=0.0358); thus we analyzed  only the first period.  

\begin{figure*}[!t]
	\centering
	\begin{subfigure}[b]{.32\linewidth}
		\includegraphics[width=\columnwidth]{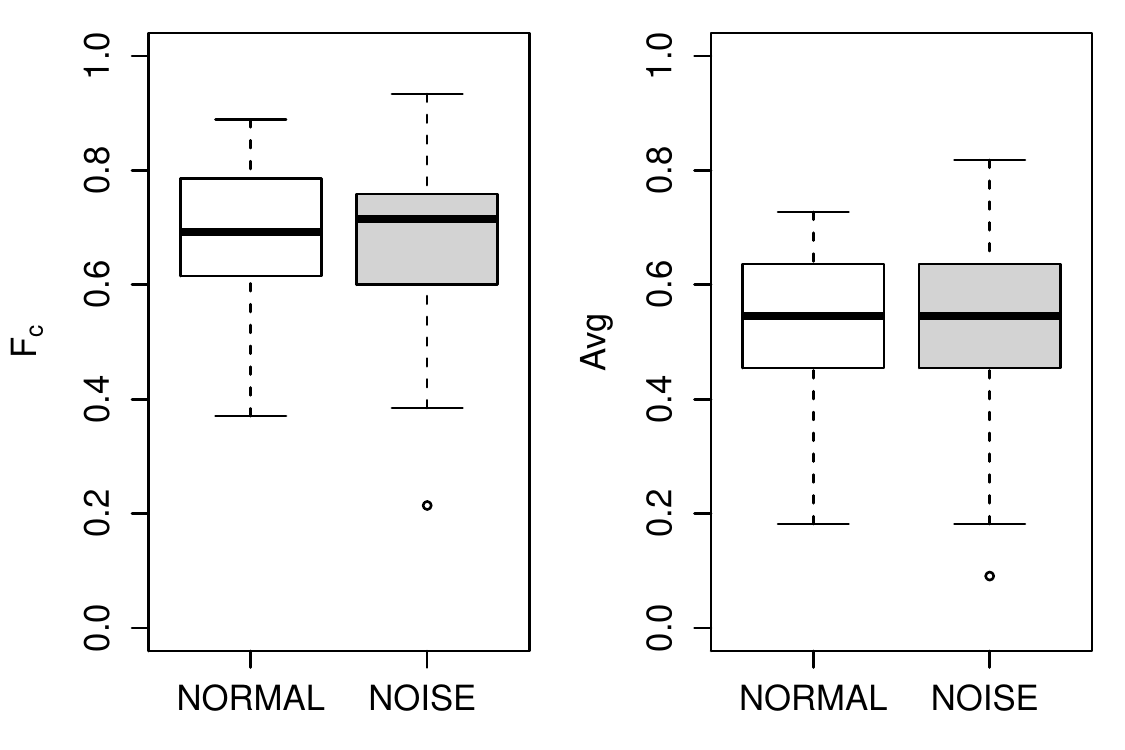}
		\caption{Exp1}
		\vspace{-0.4cm}
	\end{subfigure}
	\begin{subfigure}[b]{.32\linewidth}
		\includegraphics[width=\columnwidth]{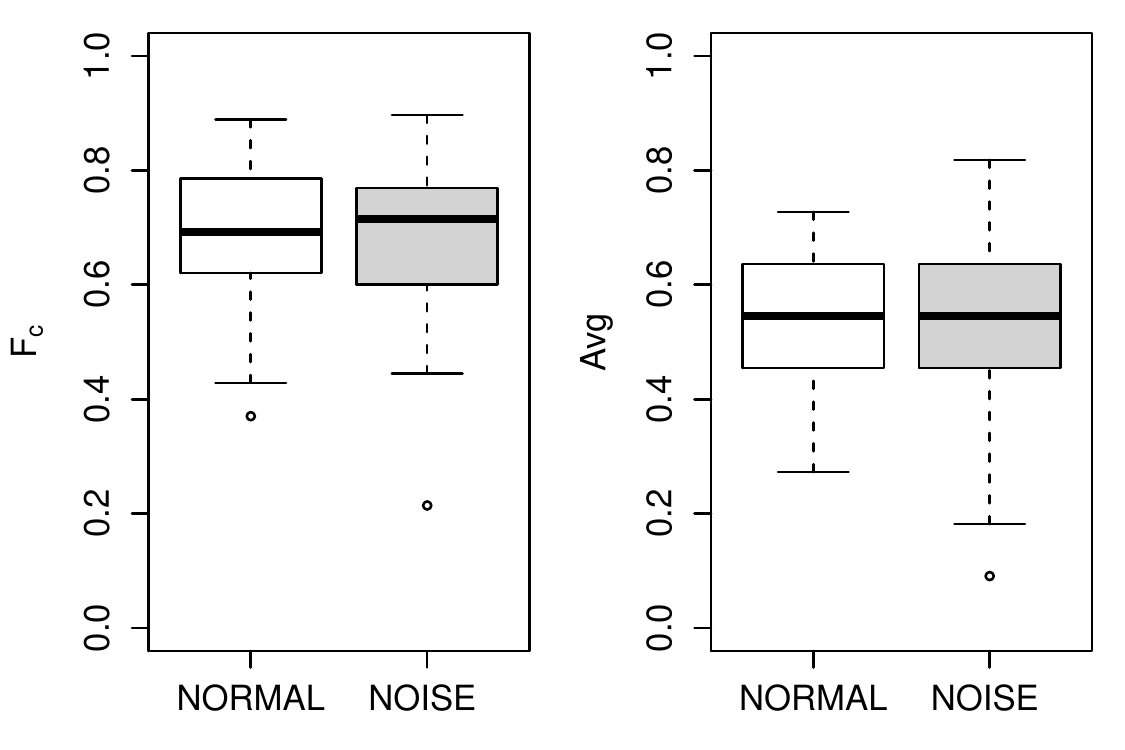}
		\caption{Exp1*}
		\vspace{-0.4cm}
	\end{subfigure}
	\begin{subfigure}[b]{.158\linewidth}
		\includegraphics[width=\columnwidth]{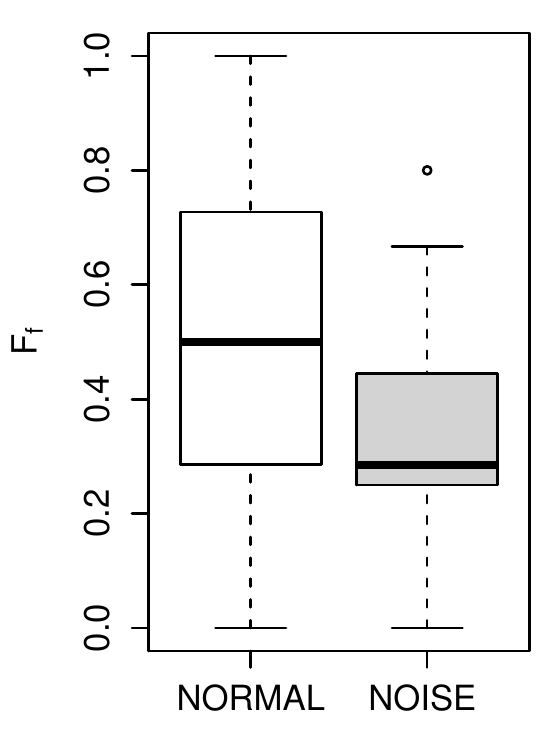}
		\caption{Exp2}
		\vspace{-0.4cm}
	\end{subfigure}	
	\caption{Boxplots for: $\boldsymbol{F_c}$ and $\boldsymbol{Avg}$ in Exp1 (a), $\boldsymbol{F_c}$ and $\boldsymbol{Avg}$ in Exp1* (b), and $\boldsymbol{F_f}$ in Exp2 (c).}
	\label{fig:boxplots}
	\vspace{-0.4cm}
\end{figure*}

\subsection{Descriptive Statistics and Boxplots}
In Table~\ref{tab:descriptive}, we summarize descriptive statistics of the dependent variable values the participants obtained in each experiment grouped by treatment (\ie NORMAL or NOISE). These values are also graphically summarized using the boxplots in Figure~\ref{fig:boxplots}. 

As for Exp1, the descriptive statistics  suggest that there is not a huge difference between the participants administered with  NORMAL and NOISE  with respect to $F_c$ (0.6781 vs 0.6833 on average) and $Avg$ (0.5025 vs 0.5256 on average). The boxplots in Figure~\ref{fig:boxplots}.a confirms this similarity. This result seems to be confirmed in Exp1*. On average, the $F_c$ values are 0.687 for NORMAL and 0.6949 for NOISE, whereas the $Avg$ values are 0.5065 and 0.5281, respectively. 

As for Exp2, the descriptive statistics of the $F_f$ values seem to indicate that there is a difference between the NORMAL and  NOISE treatments (0.4814 vs 0.3149 on average). The boxplots, shown in Figure~\ref{fig:boxplots}.c, seem to confirm this result. 

\begin{table}[t]
	\centering
	\caption{Some descriptive statistics grouped by experiment, variable, and condition.}
	\label{tab:descriptive}
	\resizebox{0.75\columnwidth}{!}{
	\begin{tabular}{lllcc}
		\hline
		Experiment& Variable &  Statistic& NORMAL & NOISE \\ \hline \hline
		\multirow{6}{*}{Exp1} & \multirow{3}{*}{$F_c$} & Median & 0.6923 & 0.7143 \\
		& & Mean & 0.6781 & 0.6833 \\
		& & Std & 0.1268 & 0.1398 \\ \cline{2-5}
		& \multirow{3}{*}{$Avg$} & Median & 0.5455 & 0.5455 \\
		&  & Mean & 0.5025 & 0.5256 \\
		& & Std  & 0.1296 & 0.1552 \\ \hline
		\multirow{6}{*}{Exp1*}       & \multirow{3}{*}{$F_c$}    & Median & 0.6923 & 0.7143 \\
		&  & Mean & 0.687 & 0.6949 \\
		& & Std & 0.118 & 0.1382 \\ \cline{2-5}
		& \multirow{3}{*}{$Avg$}    & Median & 0.5455 & 0.5455 \\
		& & Mean & 0.5065 & 0.5281 \\
		& & Std & 0.124 & 0.1646 \\ \hline
		\multirow{3}{*}{Exp2} & \multirow{3}{*}{$F_f$} & Median & 0.5 &  0.2857 \\
		& & Mean & 0.4814 & 0.3149 \\
		& & Std & 0.2969 & 0.1983 \\ \hline
	\end{tabular}
	}
\end{table}

\subsection{Hypotheses Testing}
 
\subsubsection{Hn1\textemdash comprehension of functional requirements.} 
As for Exp1, we used the t-test for $F_c$ because the data were normally distributed, while we used the Mann-Whitney U test for $Avg$ because the Shapiro test indicated that the data were not normally distributed (\ie p-value = 0.0167 for Group2). As for Exp1*, we used the t-test for both $F_c$ and $Avg$ (the Shapiro test always returned p-values greater than $\alpha$).
For both Exp1 and Exp1*, and any dependent variable, we could not reject $Hn1$ (see Table~\ref{tab:testing}).

\subsubsection{Hn2\textemdash fault fixing in source code.} 
The Mann-Whitney U test---the Shapiro test returned a p-value equal to 0.0229 for NOISE---allowed us to reject $Hn2$ with respect to $F_f$ (p-value = 0.024, see Table~\ref{tab:testing}). 
Thus, noise has a significant and negative effect on software engineers' performance in fixing faults in source code. To measure the magnitude of the difference, we used the Cliff's $\delta$ effect size~\cite{Cliff}. This kind of effect size is used when data are not normally distributed or the normality assumption is discarded. As shown in Table~\ref{tab:testing}, the effect of noise on fault fixing tasks was medium.\footnote{According to Romano \etal~\cite{romano2006appropriate},  the Cliff's $\delta$ effect size is: ``negligible'' if $|\delta|<0.147$, ``small'' if $0.147 \le |\delta| < 0.33$, ``medium'' if $0.33 \le |\delta| < 0.474$, or ``large'' otherwise.}

\begin{table}[t]
	\centering
	\caption{Results from the testing of $\boldsymbol{Hn1}$ and $\boldsymbol{Hn2}$  (p-values less than $\boldsymbol \alpha$ are reported in bold).}
	\label{tab:testing}
	\resizebox{0.8\columnwidth}{!}{
	\begin{tabular}{lllcc} \hline
		Hn & Experiment & Variable & p-value & Effect size \\ \hline \hline
		\multirow{4}{*}{Hn1} & \multirow{2}{*}{Exp1} & $F_c$ & 0.7309 & - \\ 
		& & $Avg$ & 0.3119 & - \\ \cline{2-5}
		  & \multirow{2}{*}{Exp1*} & $F_c$ & 0.7116 & - \\ 
		& & $Avg$ & 0.3782 & - \\ \hline
		Hn2 & Exp2 & $F_f$ & \textbf{0.024} & medium (0.4014) \\ \hline
	\end{tabular}
	}
\end{table}

\begin{table}[t]
	\centering
	\caption{Some descriptive statistics for $\boldsymbol{F_f}$ when tanking into account only the first 30 minutes.}
	\vspace{-0.3cm}
	\label{tab:further}
	\resizebox{0.425\columnwidth}{!}{
		\begin{tabular}{lcc}
			\hline
			Statistic & NORMAL & NOISE \\ \hline \hline
			Median & 0.2857 &  0.2857 \\
			Mean & 0.3141 & 0.1741 \\
			Std & 0.2163 & 0.2024 \\ \hline
		\end{tabular}
		
	}
\end{table}

\subsection{Further Analysis} \label{sec:furtherAnalysis}
We analyzed the data from Exp2 by taking into account the first 30 minutes (first period). This was possible because we knew when each participant started tackling any bug and when it was fixed (see Section~\ref{sec:tasks}). We perform this further analysis to exclude noise duration (Exp1 and Exp2 lasted 30 and 60 minutes, respectively) from the possible causes behind the lack of statistical significant difference in Exp1 (and Exp1*), with respect to Exp2. 

In Table~\ref{tab:further}, we report some descriptive statistics for $F_f$ with respect to the first 30 minutes of Exp2. These descriptive statistics indicate that, on average, who fixed faults in the presence of noise had worse performance than who worked in normal conditions (the mean values of $F_f$ are 0.1741 for NOISE and 0.3141 for NORMAL). To check that this difference was statistically significant, we ran a two sided Mann-Whitney U test (the data were not normally distributed for both NOISE and NORMAL as the Shapiro test suggested). The returned p-value (0.0281) indicate that the participants exposed to noise for 30 minutes had significantly worse performances in fixing fault in source code than the participants not exposed. The effect of noise was medium (the Cliff's $\delta$ effect size value was 0.3605). Thus, we can exclude that different noise durations are behind the lack of statistical significant difference in Exp1 (and Exp1*).

%% file: Discussion.tex
In the next subsections, we first discuss the results by linking them to RQ1 and RQ2. Then, we delineate practical implications from the obtained results and future directions for our research. We conclude discussing threats that could affect the validity of our results.


\subsection{Linking  Results to Research Questions and Overall Discussion}\label{sec:linkResults}

Results from Exp1 and Exp1* suggest that noise does not significantly affect the comprehension of  functional requirements. Thus, we cannot positively answer RQ1.
On the other hand, we observed that noise negatively affects  faults fixing (also restricting the analysis to the first 30 minutes in Exp2, see Section~\ref{sec:furtherAnalysis}). Thus, we can positively answer RQ2: ``noise worsens software engineers' performance in fixing faults in source~code.''
This outcome suggests that noise duration is not the cause of the lack of statistically significant difference in Exp1 (and Exp1*), but it is the kind of task. Fixing faults in source code seems to be more vulnerable to noise than comprehending functional requirements. According to the meta-analytic results by  Szalma and Hancock's~\cite{Szalma:2011}, we can then speculate that fixing faults  is more resource-demanding than comprehending functional requirements. Concluding, it seems that there are tasks in software engineering that are more resource-demanding  than others  and noise seems to negatively affect the executuion of these tasks.


\subsection{Implications and Future Directions}
We focus here on  practitioner and researcher perspectives.
\begin{itemize}[leftmargin=*]
\item Comparing our results on the comprehension of functional requirements to those reported in~\cite{Abrahao2013} (where the same experimental material and dependent variable $F_c$ were used), we observe that our participants (\ie final-years undergraduate students) performed worse than Ph.D. students (on average, 0.6781 is the $F_c$ value in Exp1, while 0.727 is the one achieved by the Ph.D. students), but better than professionals (0.6781 vs 0.631, on average). Thus, we postulate that the participants in Exp1 had an adequate level of familiarity with the used modeling notation, so allowing us to assume that they are representative of professionals. Therefore, rather than replicating the first experiment with professionals, the researcher could be interested in investigating whether different noise intensities (\ie levels)  affect software engineers' performance in comprehending functional requirements. The researcher could also be interested in varying the modeling methods (\ie UML) and assessing if the comprehension of functional requirements is still not vulnerable to noise.

\item We observed a statistically significant difference in fault fixing tasks when the participants were exposed or not to noise. This finding is relevant for the practitioner because noisy workspaces (\eg those with open-office plans) could cause a reduction in the performance of software engineers that have to fix faults in Java code. In other words, a penny saved on the workspace could be paid with interest later. However, we believe that caution is needed concerning this finding and we advise future work. For example, the researcher could investigate if Exp2 results also hold for professionals. 
Other directions for future work could consist in varying the noise intensity and type.

\item Fixing faults in source code seems to be more vulnerable to noise than comprehending functional requirements. This finding is consistent with arousal and maximal adaptability theories~\cite{Broadbent1978, Hancock1989} and the results from the Szalma and Hancock's meta-analysis~\cite{Szalma:2011}. The researcher could be interested in studying which tasks are more (or less) vulnerable to noise. That is, some kinds of tasks (\ie those less resource-demanding) could be weakly affected by noise. Thus, software companies could save money by providing workspaces with open-office plan to the software engineers involved in these kinds of tasks. On the other hand, more resource-demanding tasks should be more vulnerable to noise. Thus, quiet workspaces (\eg those with closed-office plans) are advisable. This is clearly relevant for the practitioner.

\item Given our results, we speculate that a 30-minute wash-out period is not enough as far as 1-hour fault fixing tasks is concerned. Our results provide a reference point for the duration of wash-out periods, which we did not have when we designed our experiments.  This finding is of interest for the practitioner too. For example, our results suggest software engineers to take a break
greater than 30 minutes when performing a fault fixing task in the presence of noise. Such a long wash-out period might be expensive for a software company. Therefore, it could be more advisable to provide software engineers with quieter workspaces. 

\item The presence of a significant carryover effect in Exp2 indicates a detrimental effects of noise on the performance, when fixing faults in Java code,  even when participants are no longer exposed to noise. This is relevant for the practitioner. On the other hand, the researcher could be interested in further investigating on this matter using special conceived investigations.
\end{itemize}

\subsection{Threats to Validity} \label{sec:threats}
  
We discuss threats that could affect our results by following the recommendations by Wohlin \etal~\cite{wohlin12}. 

\subsubsection{Internal Validity}
Social threats to validity might exist. The participants exposed to noise might be more motivated to accomplish the task (\eg due to arousal effect) than the participants working in normal condition~\cite{Broadbent1978,Hancock1989}. On the other hand, the participants exposed to noise might give up performing as they would do under normal conditions (\ie \textit{resentful demoralization}). This kind of threat is present  when the data analysis is conducted on the first period (\ie when dealing with a statistically significant carryover). 

The \textit{selection} threat of letting volunteers take part in the experiments could influence the results since they could be more motivated than actual developers.

To prevent that participants exchanged information on the  tasks (\ie \textit{diffusion or imitation of treatments}), we monitored them during the execution of each task and we took back all the material we gave them to accomplish the tasks. In addition, Group1 and Group2 performed the same task (\eg bug fixing on LaTazza) at the same time (see Section~\ref{sec:design}) in each experiment. 

The use of an AB/BA design might affect internal validity. We dealt with this kind of threat by means of a wash-out period in each experiment. We studied if wash-out periods were long enough to neutralize carryover effect  (Section~\ref{sec:analysisProc}). In case of carryover was present, we used the strategy suggested by Vegas \etal~\cite{Vegas:2016} (\ie taking into account only the first period in the data analysis).

\subsubsection{Construct Validity}
To deal with this kind of threat, we exploited metrics well known and adopted in the literature (\eg~\cite{Abrahao2013,ricca:tse2010,Scanniello2017}). As far as Exp2 is concerned, we considered only one metric to assess participants' performances (\ie \textit{mono-method bias}).

Although we did not inform the participants about our research goals, they were aware of being part of an investigation on noise effect. Thus, there could be the risk of \textit{hypotheses guessing}. 

We informed students that achieved performance would not affect the SE course grades and gather data would be shared anonymously and in aggregated fashion (\ie \textit{evaluation apprehension}).

\subsubsection{Conclusion Validity}

The implementation of a treatment (\eg NOISE) might be not similar between different participants (\ie \textit{reliability of treatment implementation}). As shown in Section~\ref{sec:validityConsiderations}, we mitigate this kind of threat by implementing treatment/control as standard as possible over different participants.



\subsubsection{External Validity} 
The participants were sampled by convenience.
Therefore, generalizing the results to a different population (\eg professional software developers rather than students at the University of Basilicata) poses a threat of \textit{interaction of selection and treatment}. Working with students also implies various advantages, such as: their homogeneous prior knowledge, the availability of a large number of participants~\cite{Verelst:2004} (55 and 42 participants in studies like ours is appreciable), and the possibility to test experimental design and initial hypotheses~\cite{Sjoberg:TSE:2005}. 

The use of UML as modeling notation in Exp1 might threaten the generalizability of the results (\ie \textit{interaction of setting and treatment}). 
We opted for UML because it is a de-facto standard in software modeling and the students were familiar with it. 
Another threat of interaction of setting and treatment might exist due to the non-real-world experimental tasks used. The experimental tasks should equally affect the results of the participants when exposed or not to noise. 

%% file: Conclusion.tex
We present the results of an empirical evaluation constituted of two controlled experiments. Both experiments assess whether noise negatively influences software engineers' performances. In the first experiment, we asked 55 final-year undergraduate students in Computer Science to  comprehend  functional requirements. While performing this task, the participants in this experiment were exposed or not to noise. 
We asked these participants to take part in the second experiment (42 out of 55 agreed to participate in) where the task to be performed was fixing faults in Java source code.  The results suggest that the participants had significantly worse performance in fixing faults in source code when exposed to noise, while no difference was observed in comprehending functional requirements. We conjecture that bug fixing is a more resource-demanding than comprehending functional requirements as it is more vulnerable to noise. 
Therefore, it seems that there are more resource-demanding tasks than others, and noise seems to negatively impact  tasks that are more resource-demanding.  